\documentclass[aapm,graphicx,mph,reprint]{revtex4-1}

\usepackage[mathlines]{lineno}

\usepackage{graphicx}
\usepackage{tabularx}
\usepackage[caption=false]{subfig}
\usepackage{comment}
\usepackage[percent]{overpic}
\usepackage{siunitx}
\usepackage{mathtools}
\usepackage{color}
\usepackage[percent]{overpic}

\newcounter{pcounter}
\newenvironment{pequation}{\addtocounter{equation}{-1} 
	\refstepcounter{pcounter} 
	 
	\begin{equation}}
{\end{equation}\ignorespacesafterend}

\DeclareMathOperator*{\argmin}{argmin\,}

\DeclareMathOperator{\diag}{diag}

\newcolumntype{C}[1]{>{\centering\let\newline\\\arraybackslash\hspace{0pt}}b{#1}}

\begin{document}


\title{Investigating Simulation-Based Metrics for Characterizing Linear Iterative Reconstruction in Digital Breast Tomosynthesis} 



\author{Sean D. Rose}

\author{Adrian A. Sanchez}
\author{Emil Y. Sidky}
\author{Xiaochuan Pan}
\affiliation{University of Chicago, Dept. of Radiology MC-2026, 5841 S. Maryland Avenue, Chicago IL, 60637}


\date{\today}

\begin{abstract}

\noindent\textbf{Purpose}: Simulation-based image quality metrics
are adapted and investigated for characterizing the parameter dependences
of linear iterative image reconstruction for DBT.

\par\noindent
\textbf{Methods}: Three metrics based on a 2D DBT simulation are investigated:
(1) a root-mean-square-error (RMSE) between the test phantom
and reconstructed image, (2) a gradient RMSE where the comparison
is made after taking a spatial gradient of both image and phantom,
and (3) a region-of-interest (ROI) Hotelling observer (HO) for
signal-known-exactly/background-known-exactly (SKE/BKE) 
and signal-known-exactly/background-known-statistically (SKE/BKS) detection tasks.
Two simulation studies are performed using the aforementioned
metrics, varying voxel aspect ratio and regularization strength
for two types of Tikhonov regularized least-squares optimization.
The RMSE metrics are applied to a 2D test phantom with resolution
bar patterns at varying angles, and the ROI-HO metric is applied
to two tasks relevant to DBT: lesion detection, modeled
by use of a large, low-contrast signal, and microcalcification
detection, modeled by use of a small, high-contrast signal.
The RMSE metric trends are compared with visual assessment of the
reconstructed bar-pattern phantom. The ROI-HO metric trends
are compared with 3D reconstructed images from ACR phantom
data acquired with a Hologic Selenia Dimensions DBT system.

\noindent
\textbf{Results}: Sensitivity of the image RMSE to mean pixel value
is found to limit its applicability to the assessment of DBT image
reconstruction. The image gradient RMSE is insensitive to mean pixel
value and appears to track better with subjective visualization
of the reconstructed bar-pattern phantom. The ROI-HO metric shows
an increasing trend with regularization strength for both
forms of Tikhonov-regularized least-squares; however, this
metric saturates at intermediate regularization strength indicating
a point of diminishing returns for signal detection. Visualization
with the reconstructed
ACR phantom images appear to show a similar dependence with regularization
strength.

\par\noindent
\textbf{Conclusions}: From the limited studies presented it appears that image
gradient RMSE trends correspond with visual assessment better than image RMSE
for DBT image reconstruction. The ROI-HO metric for both detection
tasks also appears to reflect visual trends in the ACR phantom reconstructions
as a function of regularization strength.  We point out, however, that
the true utility of these metrics can only be assessed after amassing more
data.

\end{abstract}


\maketitle 



%
%

%

\section{Introduction}
Image reconstruction
for digital breast tomosynthesis (DBT), whether direct or iterative, 
invariably involves a variety of parameters and implementation choices, such as voxel size, voxel aspect ratio, and regularization strength \cite{Sechopoulos2013}. 
Experience from computed tomography (CT) provides some
insight into the parameter dependence of these algorithms, but this experience alone is insufficient
for the tailoring of algorithms to DBT.  As in CT, exhaustive exploration of implementations and 
parameter settings is warranted for every algorithm, task, and DBT system design under 
consideration. Efficiently computable objective image quality metrics are necessary to facilitate 
this task.

Numerous authors have investigated metrics addressing this issue,
some quantifying image artifacts specific to DBT, \cite{Wu2004,Hu2008} some using
DBT tailored versions of traditional CT metrics, \cite{Sechopoulos2009,Saunders2009,VandeSompel2011} and others calculating task-specific
detectability indices. \cite{Park2010,Zeng2015,Reiser2010,Sanchez2015b,Richard2010,Gang2011,Young2013,VandeSompel2011}
A number of these works focus on optimization of system acquisition parameters,
\cite{Sechopoulos2009,Hu2008,Saunders2009,Reiser2010,Richard2010,Gang2011,Young2013,Park2010} or assessment of reconstruction
algorithms. \cite{Wu2004,Sanchez2015b, Zeng2015}

Metrics for DBT imaging fall into two categories: those that
are applied to actual DBT scans, and those that are applied to DBT simulations.
Metrics for real data are clearly more direct and are useful for DBT systems
characterization  and optimization, but there is the potential for error due
to the fact that the truth is unknown. In particular, for subtle features
such as small spiculations or calcifications, error in assessing their
detectability or morphology can be large. It is also precisely for these subtle
DBT features where image reconstruction parameters can have the greatest impact.
For this reason, it is also desirable to have metrics that summarize qualities
of DBT image reconstruction from simulated data.

In this work, we follow the latter approach and seek to establish
additional metrics for DBT image reconstruction
parameter characterization based on simulation, where the underlying true
object is known. To establish relevance of the proposed metrics, we compare
the metric trends with trends in the visualization of reconstructed DBT images
from simulation and actual scanner data.

Due to the limited scanning angular range, DBT cannot
provide an accurate 3D reconstructed volume.
This basic feature of DBT imaging can limit the usefulness
of image quality metrics based on tomographic image fidelity.
We illustrate this problem with studies based on image RMSE.
While absolute image fidelity is not achievable for image reconstruction
in DBT even under ideal simulation with no noise, there are features
of the scanned object which are recovered. We modify image RMSE
in such a way that the new RMSE-based measure is sensitive only
to features of the scanned object that are recoverable in a DBT scan.
We also investigate an image quality metric based on a signal detection
task. Such metrics reflect an important imaging task for DBT
and their applicability is not hindered by the inability to obtain
accurate tomographic image reconstruction in DBT. 


In this work, we investigate three image quality metrics for DBT image reconstruction, two
based on image fidelity and one based on a detection task.  The task-based metric extends on our previous
work in the context of analytic reconstruction algorithms in breast CT and DBT \cite{Sanchez2013,Sanchez2014,Sanchez2014a,Sanchez2015b,Rose2016} and constitutes a 
region of interest (ROI) based implementation of the Hotelling observer (HO). For brevity, we refer to it as an ROI-HO metric. Investigation of the image fidelity metrics extends our previous 
work on parameter selection in breast CT \cite{Rose2016} to a limited angular scanning geometry. Our studies focus on the behavior of these metrics
with respect to voxel aspect ratio and regularization strength for Tikhonov-regularized least-squares
optimization. 

The paper is organized as follows. In section \ref{sec:methods}, the reconstruction optimization problems, image quality metrics, and study designs are described. In section \ref{sec:results}
we present the results of our investigation. The results are separated into three studies: (1) a simulation study investigating the behavior of the two image fidelity metrics with varying 
voxel aspect ratio and regularization strength (2) a simulation study investigating the task-based metric with reconstructions of varying regularization strength, and (3) 
a real data study using reconstructions of the ACR mammography phantom for visual assessment of realistic reconstructions using parameters in the ranges investigated in 
the simulation studies. In section \ref{sec:discussion} we discuss our results in greater depth. Lastly, we conclude in section \ref{sec:summary}.


\begin{figure}[t!]
	\centering
	\includegraphics{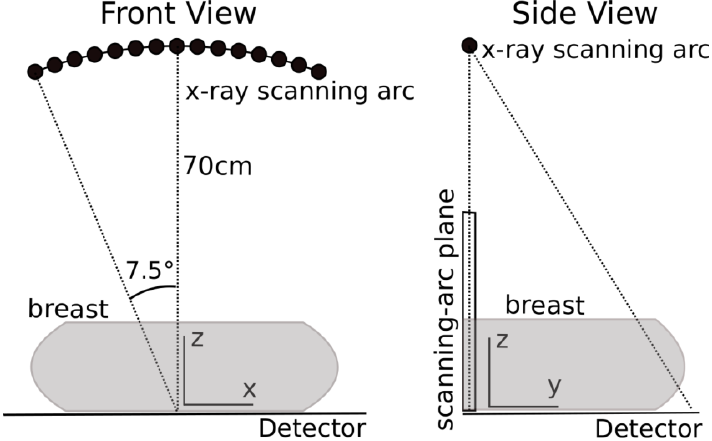}
	\caption{\label{fig:geo} Diagram illustrating system geometry and the scanning-arc plane used for 
	RMSE and the ROI-Hotelling observer calculations.}
\end{figure}

\begin{figure}[t!]
	\centering
	\includegraphics{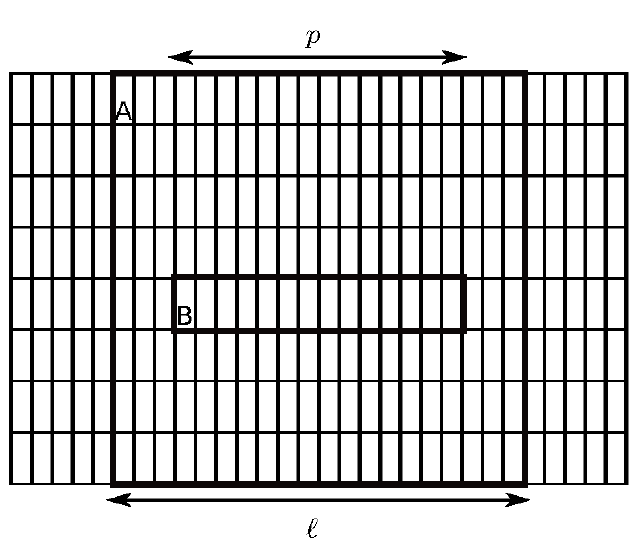}
	\caption{\label{fig:ROI} Schematic illustrating ROIs within the scanning-arc plane used in calculating the ROI-HO template. A) ROI$_\text{recon}$ used
	for calculating reconstruction operator. B) ROI$_\text{HO}$ used for computing signal detection
signal-to-noise ratio.}
\end{figure}

\section{Methods}
\label{sec:methods}

\subsection{Image Reconstruction Problems}
\label{sec:opt}
\textit{LSQI} - The first of two optimization reconstruction problems investigated in this manuscript is a least squares 
problem with identity Tikhonov regularization, referred to as LSQI. 
The optimization problem for this reconstruction formulation
can be written as
\begin{linenomath}
\begin{pequation}
	\label{eqn:LSQI}
	\argmin_x \left\{ \|A x - b\|^2 + (\lambda \|A\|)^2 \|x\|^2 \right\}
\end{pequation}
\end{linenomath}
where $x \in \Re^n$ is a pixelized image, $b \in \Re^m$ is a sinogram data vector, 
$A \in \Re^{m\times n}$ is a matrix modeling forward projection, and $\lambda \in \Re$ 
is a scalar regularization parameter. The notation $\|\cdot\|$ will be used to denote the
$\ell_2$ (Euclidean) norm throughout. When the argument of $\|\cdot\|$ is a matrix, 
it is defined to return the maximum singular 
value of its argument.

Identity Tikhonov regularization can be interpreted in 
terms of the singular value decomposition (SVD) of the system matrix $A$. Let $A = U \Sigma V^T$
be an SVD of $A$, where $\Sigma \in \Re^{m \times n}$ is a diagonal matrix with elements $\sigma_i$. 
The solution to \ref{eqn:LSQI} can be written\cite{Golub1996}
\begin{linenomath}
\begin{align*}
	x^* = V \Gamma U^T b,
\end{align*}
\end{linenomath}
where the diagonal matrix $\Gamma \in \Re^{n \times m}$ has elements
\begin{linenomath}
\begin{align*}
	\Gamma_{i i } = \frac{\sigma_i}{\sigma_i^2 + \lambda^2 \sigma_1^2}
\end{align*}
\end{linenomath}
As $\lambda \rightarrow 0$, the matrix $V \Gamma U^T$ limits to the pseudo-inverse of $A$. 
The parameter $\lambda$ can be interpreted as the fraction of the maximum singular value of $A$
below which the effects of the singular value spectrum on the reconstruction are significantly diminished.

A second important observation is that
\begin{linenomath}
\begin{align*}
	\lim_{\lambda \rightarrow \infty} \frac{x^*}{\|x^*\|} = \frac{A^T b}{\|A^T b\|}
\end{align*}
\end{linenomath}
meaning that in the limit of infinite regularization the LSQI reconstruction is visually equivalent to back-projection.


\begin{figure}[t!]
	\centering
	\includegraphics{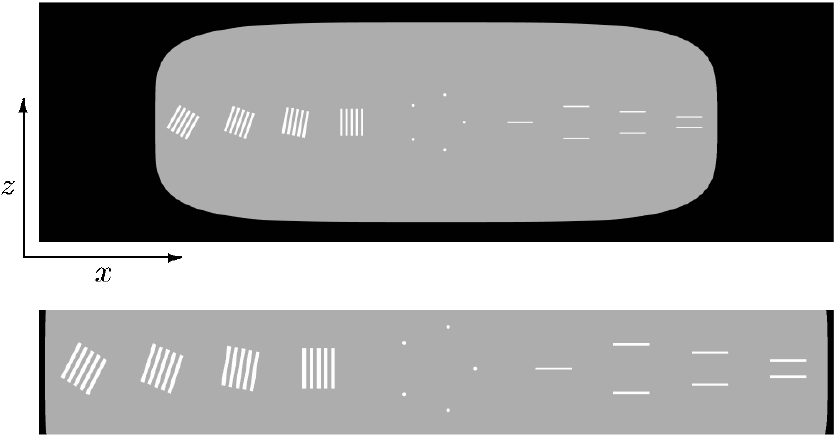}
	\caption{\label{fig:TomoPhant} Simulation phantom used for RMSE studies (top) and ROI for visualization (bottom). Coordinate axis directions are also shown. Attenuation coefficients for each
	part of the phantom were evaluated at \SI{20}{keV}. The background is fatty tissue, while the bars on the left
	and lines on the right are fibroglandular. The lines on the right are interpreted as cross sections of disks in the plane parallel to the detector in 3D. The specks in the center are calcium and are
	displayed at 5 times actual size for ease of visualization. Display window: [0.15, 0.60] \SI{}{\per\centi		
	\metre}.}
\end{figure}
\textit{LSQD} - The second problem considered is a least squares optimization with derivative Tikhonov
regularization, referred to as LSQD. The optimization problem can be written
\begin{linenomath}
\begin{pequation}
	\label{eqn:LSQD}
	\argmin_x \left\{ \|Ax - b\|^2 + (\lambda \|A\|/\|D\|)^2 \|D x\|^2 \right\}
\end{pequation}
\end{linenomath}
where $D \in \Re^{dn \times n}$ is a finite difference matrix approximation of the 
gradient operator, and $d$ is the spatial dimensionality of the image array (2-D or 3-D). The ratio 
$(\|A\|/\|D\|)$ is employed in the regularization term to normalize the strengths of the data fidelity
and regularization terms.

\begin{linenomath}
\begin{align*}
	\lim_{\lambda \rightarrow \infty} \frac{x^*}{\|x^*\|} = \frac{(D^TD)^{-1} A^T b}{\|(D^TD)^{-1} A^T b\|}
\end{align*}
\end{linenomath}
meaning the reconstruction is visually equivalent to the inverse Laplacian of the back-projection image. This limit holds if the image boundary conditions are such that boundary pixels are set to zero. In our implementation of LSQD, we enforce this boundary condition.

\textit{Linearity} - The solutions of problems \ref{eqn:LSQI} and \ref{eqn:LSQD} can be written as linear functions of the data vector $b$. Specifically, let the matrix $C$ represent the scaled linear operator involved in regularization
\begin{linenomath}
\begin{align*}
C = \begin{cases} \lambda \|A\| I &\text{ if LSQI} \\
			    \left(\lambda \|A\|/\|D\| \right) D &\text{ if LSQD}
	\end{cases}
\end{align*}
\end{linenomath}
Formulate the matrix $\tilde{A}$ and the vector $\tilde{b}$
as
\begin{linenomath}
\begin{align}
	\label{eqn:Atilde}
	\tilde{A} \coloneqq \begin{pmatrix} A \\ C \end{pmatrix} ; \quad \tilde{b} \coloneqq \begin{pmatrix} b \\ 0 \end{pmatrix}
\end{align}
\end{linenomath}
Then the solution to each reconstruction problem can be written
\begin{linenomath}
\begin{align*}
	x^* = \tilde{A}^\dagger \tilde{b}
\end{align*}
\end{linenomath}
where $\tilde{A}^{\dagger}$ is the pseudo-inverse of $\tilde{A}$. Since only the first $m$ elements of $\tilde{b}$ are non-zero, we can 
simplify this expression by writing $\tilde{A}^\dagger$ in block-form
\begin{linenomath}
\begin{align*}
	\tilde{A}^\dagger = \left( R \quad H \right)
\end{align*}
\end{linenomath}
where $R \in \Re^{n\times m}$ and $H \in \Re^{n \times n}$ for LSQI or $H \in \Re^{n \times dn}$ for LSQD, so that
\begin{linenomath}
\begin{align*}
	x^* = R b
\end{align*}
\end{linenomath} 

\textit{Implementation} - The conjugate gradient least squares (CGLS) (ref. \onlinecite[p.289]{Bjorck1996}) algorithm was used for reconstruction for both LSQI and LSQD.
The algorithm was run to numerical convergence in all cases. Direct inversion was employed in implementing the ROI-HO metric. Further details will be provided
later in the manuscript when the ROI-HO metric is discussed in greater detail. The matrix $A$ was defined using a distance driven forward projection
model.\cite{DeMan2004}

\textit{FBP} - In previous work we investigated parameter selection for filtered back-projection (FBP) reconstruction with a Hanning apodizing window in DBT. \cite{Sanchez2015b} Here we include FBP results as a reference for comparison in simulation studies. The cutoff frequency parameter for the Hanning window is specified as a fraction of the Nyquist frequency
\begin{linenomath}
\begin{align*}
	f_{\textit{max}} = \frac{1}{2 \Delta u}
\end{align*}
\end{linenomath}
where $\Delta u $ is the detector bin size. Note that decreasing the cutoff frequency in FBP reconstruction is analogous, though not equivalent, to increasing regularization strength in LSQI and LSQD reconstruction in the sense that it reduces sensitivity of the reconstruction to high-frequency components in the data. For this reason, we will refer to decreasing the Hanning window cutoff frequency as increasing the regularization strength.

In the limit of infinite regularization, the Hanning window becomes an impulse at zero frequency. In our implementation of FBP, as is common, the zero frequency value of the ramp filter is actually a small nonzero number (ref. \onlinecite[p. 74]{Kak1988}). In the limit of infinite regularization, each projection is therefore filtered by a scaled impulse at zero frequency in the Fourier domain. This is equivalent to convolving with a constant function in the spatial domain. Each filtered projection is a constant as a function of detector bin, and the reconstruction is the backprojection of these ``flat" projections.

\subsection{Simulation-based Image Quality Metrics}
\label{sec:simmets}
Three image quality metrics were investigated in our studies: (1) image RMSE, (2) gradient image RMSE, and (3)
an ROI-based implementation of the HO. For each of the three image quality metrics we restrict the DBT system model to a 2D plane perpendicular to the detector containing the X-ray source trajectory which we refer to throughout as the scanning-arc plane, illustrated in Figure \ref{fig:geo}. 
The idea of using the scanning-arc plane for computing the simulation-based image quality metrics
is that the principle imaging properties of DBT can be captured in this plane. Use of this 2D-model
also significantly decreases computation time facilitating parameter-dependence studies for accurate
solution of LSQI and LSQD.


\textit{Image RMSE} - Given an image vector $\tilde{x} \in \Re^n$ and a reconstruction $x \in \Re^n$, the image RMSE is defined as
\begin{linenomath}
\begin{align*}
	\mathrm{RMSE} =  \frac{\|x- \tilde{x}\|}{\sqrt{n}}
\end{align*}
\end{linenomath}
This global image fidelity metric is a normalized version of the Euclidean distance between two image vectors. It is sensitive to the mean pixel values of the image, and in DBT, since the reconstruction is only quasi-3D, the mean pixel values of the reconstruction can be quite
different than those of the true image.

\textit{Gradient Image RMSE} - In an attempt to obtain a metric with less sensitivity to the mean  pixel values of reconstructions, we study the 
gradient image RMSE, defined as 
\begin{linenomath}
\begin{align*}
	\mathrm{gradient \, RMSE} = \frac{\|D(x-\tilde{x})\|}{\sqrt{n}}
\end{align*}
\end{linenomath}
where, as before, $D \in \Re^{d\times n}$ is a finite differencing operator. This metric can be interpreted as the summed RMSE of the $d$ gradient 
images. 
\begin{figure}[t!]
	\centering
	\includegraphics{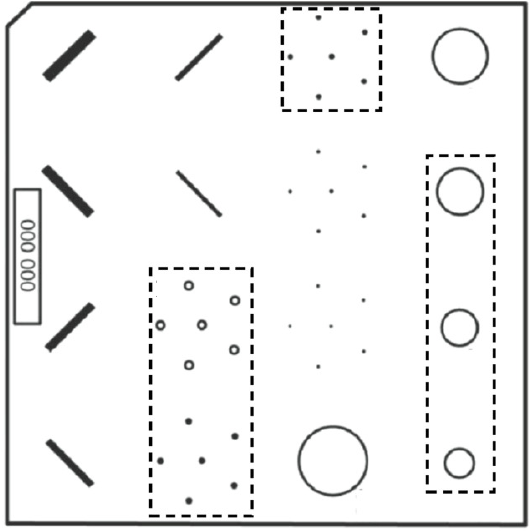}
	\caption{\label{fig:ACRDiagram} Diagram of ACR phantom. Dashed ROIs are used for
	visualization in the ACR study.}
\end{figure}

\textit{ROI-HO Metric} - The last metric we investigate 
focuses on how effectively a given task can be performed using the information
within a specified ROI of a reconstruction. The figure of merit for a detection task
performed by the ROI-HO is a signal-to-noise ratio (SNR), which quantifies
the ability of the ROI-HO to classify a given image into one of two cases: signal-present,
i.e. signal plus background, or
signal-absent, i.e. background only. 
We apply the ROI-HO metric to two different types of detection tasks: SKE/BKE
and SKE/BKS. 

For the SKE/BKE detection task, the confounding physical factor is
quantum noise, and the noise model is taken to be additive, zero-mean Gaussian with
variance proportional to the noise-free projection of the background object.
The signal is assumed small, and we employ the same noise model for signal-present
and signal-absent cases.

For the SKE/BKS detection task, in addition to quantum noise, the presence of random nonuniform background structure 
is considered, further complicating signal detection. A statistical distribution is used to model this
background variability. The model used here assumes the background structure can be described by a nonuniform component, modeled
as a zero-mean stationary random process, added to a uniform background component of known attenuation.\cite{Myers1990} 
A radially symmetric noise power spectrum of the form
\begin{align}
	\label{eqn:nps}
	h_{bg}(\nu) = \alpha |\nu|^{-\beta}
\end{align}
is considered for the nonuniform component,\cite{Metheany2008,Chen2012,Chen2013} with $\beta=1.581$.\cite{Vedantham2012} The value of $\alpha$ is chosen so that the full width at half maximum of each pixel's 
marginal distribution is equal to the difference between attenuation coefficients of fibroglandular and fatty breast tissue at 20keV.

The SNR for signal detection in the image is given by
\begin{linenomath}
\begin{align*}
	\mathrm{SNR}^2_x = w_x^T  s_x
\end{align*}
\end{linenomath}
where $w_x$ is the image domain Hotelling template, and $s_x$
is reconstruction of the signal of interest. The Hotelling template is defined
implicitly through the equation
\begin{linenomath}
\begin{align}
	\label{eqn:template}
	K_x w_x = s_x
\end{align}
\end{linenomath}
where $K_x$ is the reconstructed image covariance matrix.

The ROI-HO is a form of the Channelized Hotelling Observer where the channels are the 
pixels within an ROI.  In defining the ROI-HO, we perform a decimation 
operation $M \in \Re^{p \times n}$ which removes pixels outside of a specified ROI. The ROI is referred to as ROI$_\text{HO}$ and
is illustrated schematically (ROI sizes in the Figure are not reflective of what was used in experiments)  in Figure \ref{fig:ROI}. This
restricts the observer to a 0.42cm region (30 pixel width)  of a single row of the reconstructed image containing the true
volume of the physical signal. The resulting task is analogous to the case in 3D DBT image reconstruction of determining
whether or not a signal is present from an ROI within a single axial slice of the reconstruction.

Denoting the linear reconstruction operator by $R \in \Re^{n \times m}$, the image covariance matrix after decimation can be expressed
\begin{linenomath}
\begin{align}
	\label{eqn:cov}
	K_x = M R K_b R^T M^T
\end{align}
\end{linenomath}
where $K_b \in \Re^{m\times m}$ is the data covariance matrix. Due to the low dimensionality of $K_x$ resulting from the decimation operation, direct inversion of the
matrix can be performed for calculation of the Hotelling template.

While decimation conveniently allows for efficient direct inversion of $K_x$, calculation of the reconstruction matrix $R$ still poses a problem. Doing so directly requires pseudo-inversion of the concatenated matrix $\tilde{A}$, defined in equation \ref{eqn:Atilde}. The large dimensionality of $\tilde{A}$ makes this
a time consuming step. The idea of exploiting local shift invariance may help with this issue. \cite{Fessler1996,Qi2000}  For this work, however, we calculate an approximation of $R$ using a subset of image pixels and detector bins.\cite{Cloquet2013,Sanchez2014a} In particular, in calculating the forward projection
operator $A$ we consider only the middle $l > p$ columns of the image and, correspondingly, only the middle $l$ detector bins in each projection. This is illustrated by ROI$_\text{recon}$ in figure 
\ref{fig:ROI}. The evaluation of the regularization
matrix $C$ is similarly confined in the image space. This defines a matrix $A^{(l)} \in \Re^{m_l \times n_l}$ and a corresponding matrix $\tilde{A}^{(l)}$ of low
enough dimensionality for efficient direct inversion, yielding a reconstruction operator $R^{(l)} \in \Re^{n_l \times m_l}$. The scalars $m_l$ and $n_l$ refer to the number of measurements
and image pixels when considering only the middle $l$ detector bins and image columns, respectively. 

Referring to equation \ref{eqn:cov}, the question of how well this approximation holds is dependent on how well the matrix $R^{(l)} K_b^{(l)} R^{(l) T}$ approximates the elements of $R K_b R^T$ 
within the middle $l$ columns of the image. As the reconstruction operator $R^{(l)}$ can vary significantly with the size of $l$, it is important to check 
that $l$ is large enough for the approximation to be reasonable. This was done by calculating the ROI-HO metric for a range of values of $l$.

Instead of reporting SNR$_x$ directly, we report the efficiency $\epsilon$ which relates SNR$_x$ to
the HO SNR for signal detection in the projection data domain
\begin{linenomath}
\begin{align*}
	\mathrm{SNR}^2_b = s^T_b (K_b^{-1} s_b),
\end{align*}
\end{linenomath}
where $ s_b$ is the projection of the signal of interest. The expression in the parenthesis is
the data domain Hotelling template.


The efficiency \cite{Tanner1958} is then written as
\begin{linenomath}
\begin{align*}
	\epsilon = \frac{\mathrm{SNR}^2_x}{\mathrm{SNR}^2_b}.
\end{align*}
\end{linenomath}
The efficiency is less than or equal to one for linear image reconstruction
and it is a measure of how well the relevant information for performing a given task was preserved through the reconstruction and decimation procedure.

In order to compare detection performance for SKE/BKE and SKE/BKS tasks directly, we will report the relative efficiency for SKE/BKS tasks, which we define as 
\begin{align*}
	\epsilon_r = \frac{\mathrm{SNR}^2_{x,bg}}{\mathrm{SNR}^2_b}.
\end{align*}
where $\mathrm{SNR}^2_{x,bg}$ is the image domain ROI-HO SNR for an SKE/BKS task and $\mathrm{SNR}_b$ is the data domain HO SNR for an SKE/BKE task. The relative efficiency $\epsilon_r$ for a SKE/BKS task will always be less than the efficiency $\epsilon$ for the corresponding SKE/BKE task. Due to the common normalization factor, comparison of the SKE/BKE efficiency $\epsilon$ with the SKE/BKS relative efficiency $\epsilon_r$ is equivalent to comparing their image domain squared SNRs.

\textit{Subjective visualization} - Along with plots of the image quality metrics, we show reconstructed images corresponding to some investigated parameter settings to allow for comparison 
of metrics with subjective visualization of the 2D-model DBT reconstructions. In addition, we also show 3D DBT images reconstructed from phantom data acquired with a DBT scanner.

\subsection{DBT simulation studies}
We present two specific DBT simulation studies that characterize parameter dependences of image
reconstruction by LSQI and LSQD. The simulation studies have two purposes: (1) to demonstrate applicability of the presented image quality metrics, and in the process, (2) to reveal important characteristics of linear iterative image reconstruction for DBT.
The simulation studies are not meant to be comprehensive as the main purpose is the development
and demonstration of the image quality metrics.

\textit{Simulation System Geometry} -
For the real data studies, projection data was acquired using
a Hologic Selenia Dimensions unit. For simulation studies, the 2D system geometry was defined to 
mimic the geometry of the measurements taken by a single detector row of the Hologic system. Specifically, the radius of the source trajectory and source-to-detector
distance were both taken to be \SI{70}{\centi\metre} and 15 simulated projections were acquired in 1 degree increments. The detector
bin size was taken to be \SI{0.14}{\milli\metre}.

\subsubsection{Simulation study 1: Pixel anisotropy and regularization strength}
\label{sec:sim1}
The first study aims at characterizing image reconstruction by LSQI and LSQD as
a function of pixel anisotropy and regularization strength for a pseudo-continuous model of mean DBT data.
The dominant feature of the DBT scanning system is its limited scanning angular range.
This causes the inherent resolution of the DBT system to be highly anisotropic. Accordingly, this first
study focuses on setting the pixel aspect ratio and the impact of regularization, isolating the limited    scanning angular range aspect of the DBT system. The idea is to characterize what features of the test object can be recovered from the DBT sampling conditions and how well the RMSE-based metrics reflect the recovery of visual features. 

For this first study, a discrete phantom is defined on a high-resolution grid with pixels much smaller than the detector bins so that its projection yields pseudo-continuous data. The pixel grid is $4096 \times 4096$ with
of $x \times z$ dimensions $\SI{20.5}{\centi\metre} \times \SI{4.5}{\centi\metre}$ and
pixel dimensions of $\SI{50}{\micro\metre} \times \SI{11}{\micro\metre}$. An image of the phantom is shown in 
Figure \ref{fig:TomoPhant} along with the axis directions. The $x$ axis lies parallel to the detector plane.
The bars on the left provide visual assessment of resolution at small angles from the horizontal axis. The disks on the right are meant 
for both evaluating detection of large, low contrast signals and assessing resolution in the 
vertical direction.
This phantom is defined in the scanning-arc plane and its use for visualization is non-standard,
because it is oriented perpendicular to the usual 3D DBT slice images.  
Visualization of reconstructions in this non-standard orientation is simplified by the phantom's uncomplicated design.

Data was generated by projecting the phantom onto the detector line at 100 times the detector bin resolution using the line-intersection method. This high-resolution projection was subsequently converted to X-ray transmission employing Beer's law, assuming monochromatic X-rays at 20 keV. The transmission factors were then averaged within each detector bin, then converted back to  projection data by taking the negative logarithm.
Reconstruction was performed onto grids covering the same region as the high resolution grid on which the phantom
was defined.
The $x$ pixel dimension of reconstructions was held constant at 0.14mm --- matched to the detector bin size --- 
while the $z$ pixel dimension was varied.

The image RMSE was calculated by first resampling, via nearest-neighbor interpolation, the reconstructed image to the high-resolution grid, then calculating the RMSE. 
The gradient RMSE was evaluated by calculating a finite difference approximation of the gradient at the reconstructed resolution
then resampling the $x$ and $z$ direction gradient images to the high resolution grid. 
Resampled gradient images were compared to the gradient images of the 
phantom to calculate the gradient image RMSE.
Both metrics were plotted as a function
of the pixel aspect ratio, defined as the ratio of $z$ to $x$ pixel dimension.
This process was repeated to obtain RMSE curves at different regularization strengths. For reference, FBP reconstructions were performed at an aspect ratio of 1.0. This aspect ratio for FBP was picked because it yielded the minimum gradient image RMSE.

\subsubsection{Simulation study 2: ROI-HO detection efficiency as a function of regularization strength}
\label{sec:HOMethod}

The second study improves on the data model realism of the first study by including quantum noise
and characterizes image reconstruction as a function of regularization strength for LSQI and LSQD.
The impact of the regularization strength is quantified by the ROI-HO detection efficiency $\epsilon$. 

We investigate two types of detection tasks as defined in section \ref{sec:simmets}: SKE/BKE and SKE/BKS. For SKE/BKE tasks, observer performance is purely quantum limited,
and the task does not challenge the reconstruction to resolve the signal of interest from other nearby structures which can differ between data sets. An SKE/BKE task only challenges the reconstruction to keep the signal distinguishable from the quantum noise. An SKE/BKS task with nonuniform background further challenges the reconstruction to resolve the signal from variable background structure. In DBT, background structure is of particular importance due to blur in the depth direction.


Two signals were considered for investigation in both SKE/BKE and SKE/BKS detection tasks. The first signal was a \SI{0.25}{\centi\metre} disk lying
parallel to the detector plane with X-ray attenuation coefficient equal to that of fibroglandular breast tissue at 20keV. As the study was performed in 2D with a slice perpendicular to the detector plane, a cross section of the disk --- i.e. a rectangle --- was used as the signal. The second signal was 
a \SI{0.32}{\milli\metre} micro-calcification modeled as a Gaussian with full width at half maximum (FWHM) equal to the calcification's width.


For the SKE/BKE tasks, a uniform \SI{4.2}{\centi\metre} $\times$ \SI{4.2}{\centi\metre}  background was considered 
using the attenuation coefficient of fatty tissue at \SI{20}{keV}. For the SKE/BKS tasks, the average of the attenuation coefficients of fatty and fibroglandular tissues at \SI{20}{keV} was used for the known uniform component of the background.

The size of ROI$_{\mathrm{HO}}$ was held at $p=30$ throughout (see Figure \ref{fig:ROI}), while the size of ROI$_\mathrm{recon}$
was varied from 40 to 70 or 70 to 100 for LSQI and LSQD reconstruction, respectively. The $x$ pixel dimension of reconstructions was \SI{0.14}{\milli\metre} and the aspect ratio was 9.2. 


The quantum noise model is based on a Gaussian approximation to Poisson noise. The variance of the sinogram data for the SKE/BKE tasks is (ref. \onlinecite[p. 542]{Barrett1996})
\begin{linenomath}
\begin{align*}
	\mathrm{Var}(b_i) = \frac{1}{\bar{N}_i} + \frac{1}{\bar{N}_0}
\end{align*}
\end{linenomath}
where $\bar{N}_0$ denotes the average number of incident X-rays and $\bar{N}_i$ denotes the mean number
of detected photons in the $i^{th}$ ray after passing through the object.
The mean X-ray transmission is
\begin{linenomath}
\begin{align*}
\bar{N}_i = \bar{N}_0 \exp ( - \bar{b}_i ),
\end{align*}
\end{linenomath}
where the mean projection is computed in the same way described in Sec. \ref{sec:sim1}.
The data was assumed to be a vector of identically 
independently distributed (i.i.d.) Gaussian random variables, and accordingly, off-diagonal elements of the covariance are zero. Explicitly, the data domain covariance for 
the SKE/BKE task is 
\begin{linenomath}
\begin{align}
	\label{eqn:NoiseModel}
	(K_{b,\mathrm{BKE}})_{ij} = \begin{cases}  \frac{1}{\bar{N}_i} + \frac{1}{\bar{N}_0} \quad & \text{if} \quad i=j \\
						 0 \quad &\text{if} \quad i \neq j \end{cases}.
\end{align}
\end{linenomath}

To calculate 
$K_b$ for the SKE/BKS task, the image domain background noise model given in equation \ref{eqn:nps} is propagated through the projection model and added
to the quantum noise via the equation
\begin{align*}
	K_{b,\mathrm{BKS}} = A K_{bg} A^T + K_{b,\mathrm{BKE}}
\end{align*}
where $K_{bg}$ is the covariance matrix of the variable background. The fact that the contributions of the quantum noise and background variability to the data covariance 
matrix are additive has been demonstrated in ref. \onlinecite{Myers1990}.
The image covariance matrix $K_{bg}$ is calculated from the variable background noise power spectrum via the equation
\begin{align*}
	K_{bg} = W^* \diag(u) W
\end{align*}
where $W$ is the matrix representation of the discrete Fourier transform (DFT), $u$ is a vector with elements equal to discrete samples of $h_{bg}$, and $\diag(u)$ is 
a diagonal matrix with diagonal elements $u$.

\begin{figure}[t!]
	\centering
	\includegraphics{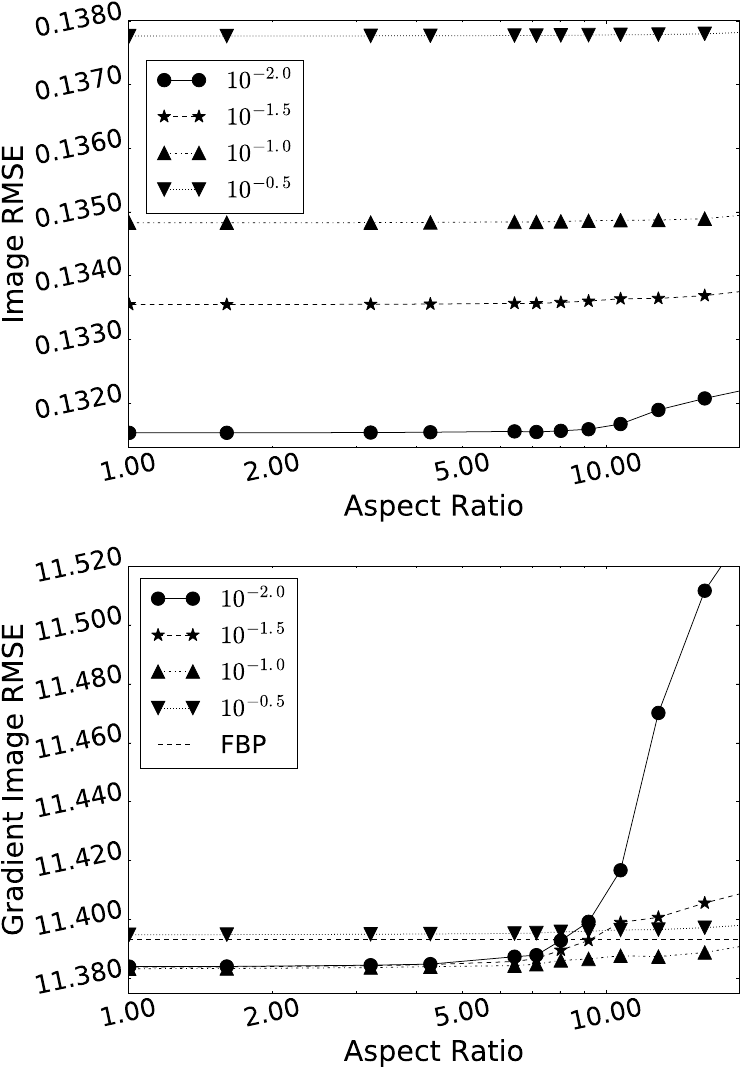}
	\caption{\label{fig:TikRMSE} Image RMSE (top) and gradient image RMSE (bottom) as a function of relative pixel size at four regularization strengths
	for LSQI reconstruction. Legends indicate value of regularization parameter $\lambda$.}
\end{figure}

\subsection{Real data study: ACR phantom data}
\label{sec:ACRmethod}
In order to show the potential relevance of the 2D ROI-HO metric to the actual 3D DBT system we
present 3D DBT reconstructed images by LSQI and LSQD for physical phantom data.  
Projection data were acquired for the mammography accreditation phantom listed by the American College of Radiology (ACR) for measuring the physical standards baseline in mammography since the beginning of the Mammographic Quality Standards Act (MQSA).
Reconstruction was performed onto a grid of pixels with $x \times y \times z$ pixel dimensions $\SI{0.14}{\milli\metre}\times \SI{0.14}{\milli\metre} \times \SI{1.3}{\milli\metre}$,
where $z$ is the direction perpendicular to the detector plane. Note the same $x$ pixel dimension was used in the simulation studies.
The three ROIs indicated in Figure \ref{fig:ACRDiagram} were chosen for visual evaluation of reconstructions.

To obtain reconstructions more representative of the 2D SKE/BKS detection tasks, a realization of nonuniform background structure was simulated, projected, and added to the measured data prior to reconstruction. The realization of nonuniform background was simulated by filtering white noise with the square root of $h_{bg}$ in equation \ref{eqn:nps} as described in detail in ref. \onlinecite{Burgess2007}.



\begin{figure}[t!]
	\centering
	\includegraphics{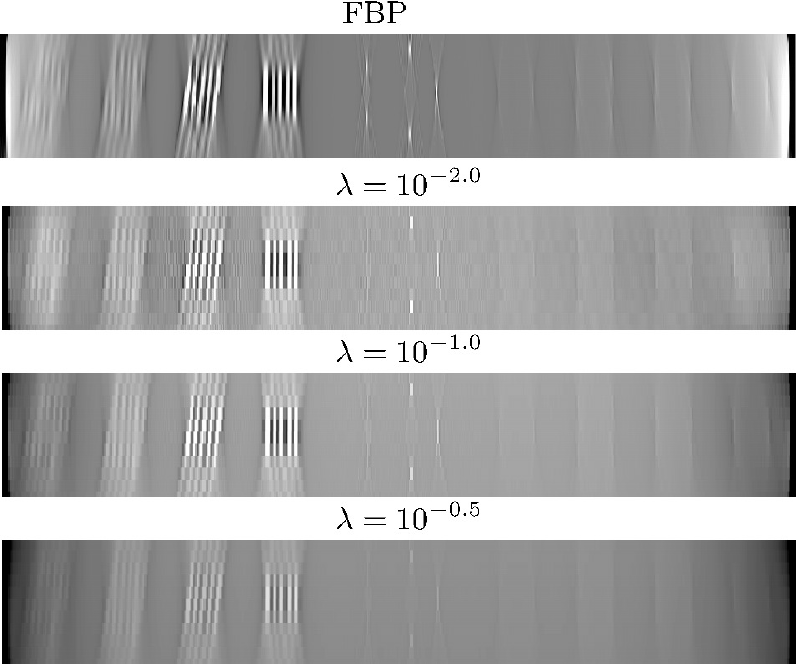}
	\caption{\label{fig:TikCond} FBP reference and LSQI reconstructions at three different regularization strengths at an aspect ratio of 10.7.
	LSQI display window: [0.15, 0.55] \SI{}{\per\centi\metre}. FBP display window: [-0.05,0.06] \SI{}{\per\centi\metre}}
\end{figure}


\begin{figure}[t!]
	\centering
	\includegraphics{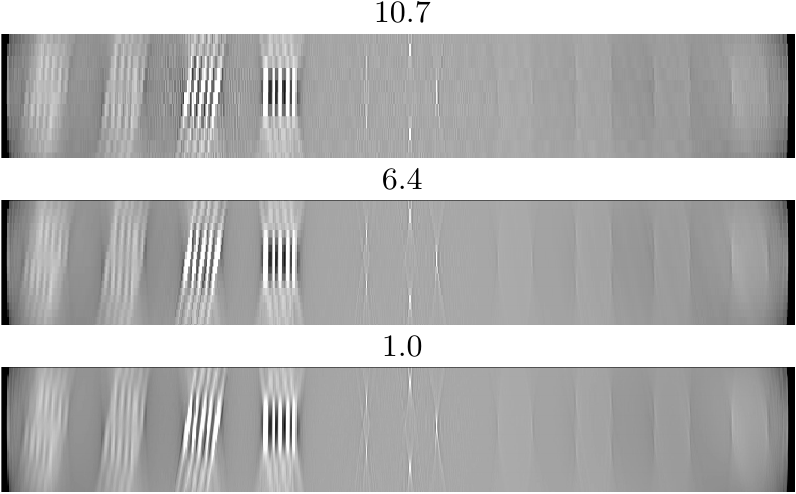}
	\caption{\label{fig:TikPix} LSQI reconstructions at $\lambda = 10^{-2.0}$ and three aspect ratios. Aspect ratios
	are displayed above each image. Display window:
	[0.15, 0.55] \SI{}{\per\centi\metre}.}
\end{figure}


\begin{figure}[t!]
	\centering
	\includegraphics{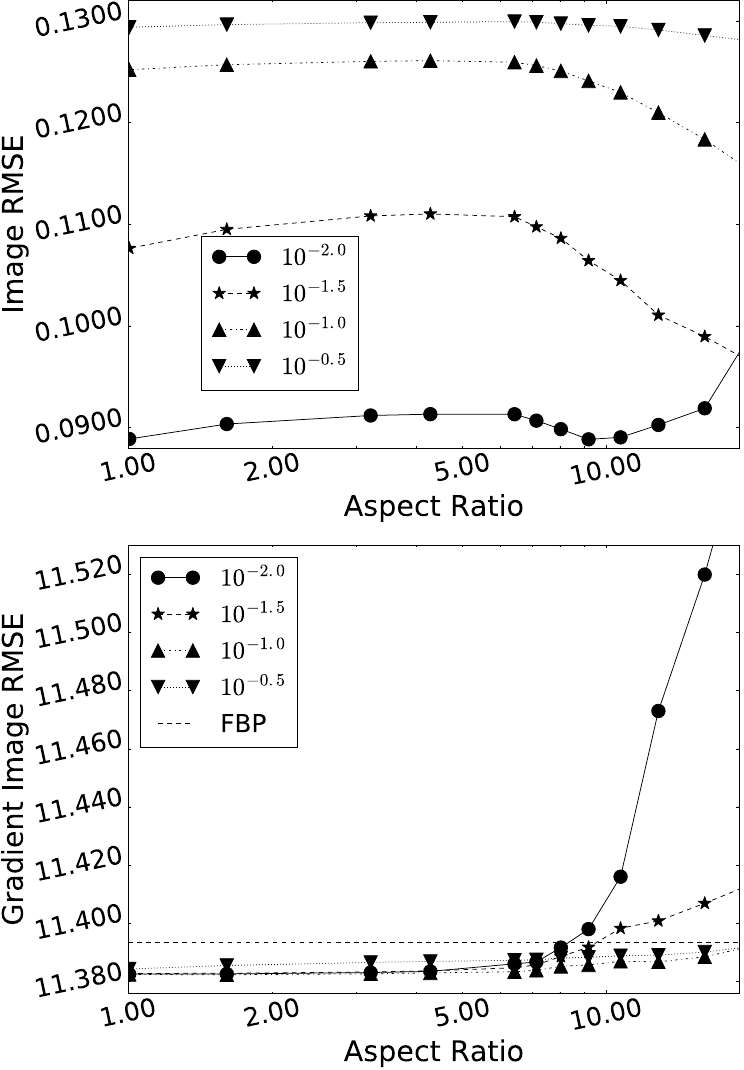}
	\caption{\label{fig:DTikRMSE} Image RMSE (top) and gradient image RMSE (bottom) as a function of relative pixel size at four regularization strengths
	for LSQD reconstruction. Legends indicate value of regularization parameter $\lambda$.}
\end{figure}


\begin{figure}[t!]
	\centering
	\includegraphics{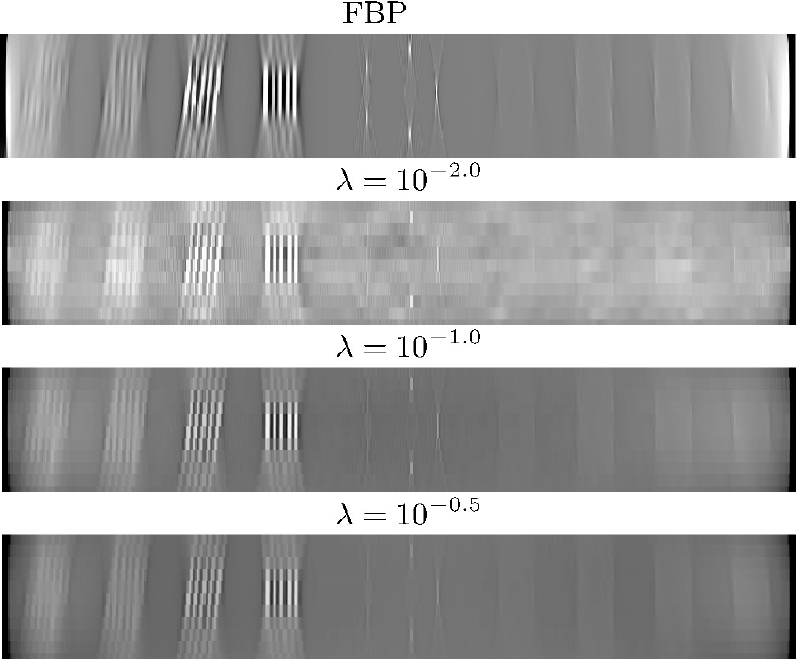}
	\caption{\label{fig:DTikCond} LSQD reconstructions at three different regularization strengths at an aspect ratio of 10.7.
	Display window: [0.15, 0.65] \SI{}{\per\centi\metre}.  FBP display window: [-0.05,0.06] \SI{}{\per\centi\metre}}
\end{figure}


\begin{figure}[t!]
	\centering
	\includegraphics{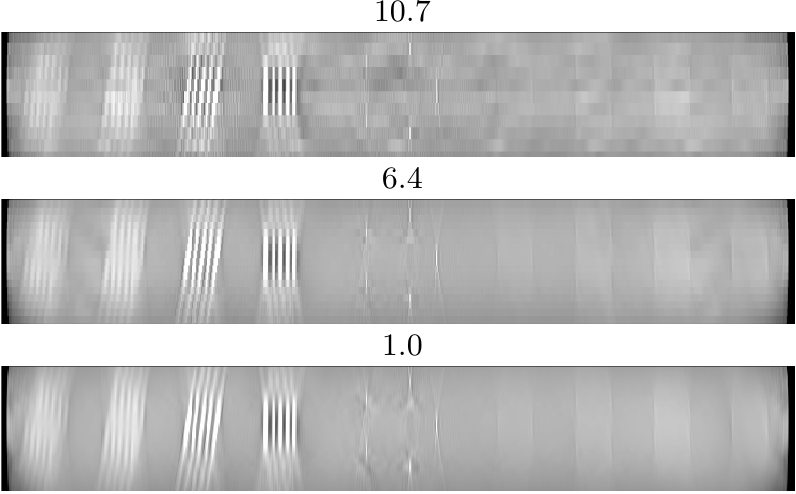}
	\caption{\label{fig:DTikPix} LSQD reconstructions at $\lambda = 10^{-2.0}$ and three aspect ratios. Aspect ratios are displayed above
	each image. Display window:
	[0.15, 0.65] \SI{}{\per\centi\metre}}
\end{figure}


\section{Results}
\label{sec:results}
\subsection{Simulation study 1: Pixel anisotropy and regularization strength}
\textit{LSQI} - Image and gradient image RMSE curves from LSQI reconstructions are shown in Figure \ref{fig:TikRMSE}. The
image RMSE of the FBP reconstruction is not shown in the top panel because it lies at value of 0.289, far above the RMSE of the LSQI reconstructions. This results from 
the FBP reconstruction having almost no DC component. Image RMSE is seen to be relatively insensitive to aspect ratio while increasing with increasing regularization strength. 
The gradient image RMSE shows a decreasing trend with decreasing aspect ratio for each level of regularization, noting that the curves for $\lambda \le 10^{-1.0}$ exhibit more of a plateau for aspect ratios $\le 5.0$.
The rank order of the curves appears to switch rapidly for aspect ratios in the range of $8.0$ to $10.0$, and
the sensitivity of this metric with regularization strength appears to increase with aspect ratio during and beyond this critical range. We note that, for all aspect ratios, image RMSE favors the smallest shown regularization strength of $\lambda=10^{-2.0}$, while gradient image RMSE favors $\lambda=10^{-1.0}$.  

The image dependence of the regularization strength is seen in Figure \ref{fig:TikCond} in which reconstructions at a fixed aspect ratio of 10.7 are shown. This aspect ratio is chosen for visualization because both RMSE measures show sensitivity to regularization strength at 10.7. At the lowest shown regularization strength $\lambda = 10^{-2.0}$, there are clear high frequency artifacts. These artifacts decrease in conspicuity with increasing regularization.  The reconstructed images also show increased blur and decreased mean value as regularization strength increases.
We also note that the FBP image appear to be of comparable image quality to the LSQI images.
Image RMSE ranks these images in the order of best to worst as $\lambda = 10^{-2.0}$, $10^{-1.0}$, $10^{-0.5}$, and FBP — with FBP being ``off-the-charts" bad. Gradient image RMSE ranks the images in the order
$\lambda = 10^{-1.0}$, FBP, $\lambda=10^{-0.5}$, and $10^{-2.0}$. The latter ordering of gradient image RMSE appears
to be more in line with subjective visualization of the images in Figure \ref{fig:TikCond}.

In Figure \ref{fig:TikPix}, reconstructions at three aspect ratios are shown at a constant level of regularization for $\lambda= 10^{-2.0}$. This value is chosen because image RMSE dependence of aspect ratio is nearly constant, while gradient image RMSE shows much variation particularly for large aspect ratios. 
Again, subjective visualization agrees more with the gradient image RMSE trend.

\textit{LSQD} - Image and gradient image RMSE curves for LSQD reconstruction are shown in Figure \ref{fig:DTikRMSE}. The trends of these curves are quite similar to those of LSQI shown in Figure \ref{fig:TikRMSE}.  The main difference is that the image RMSE curve for LSQD does actually show
some non-trivial dependence on aspect ratio as opposed to the nearly constant dependence shown for LSQI.
 Again, the FBP image RMSE value is not indicated in Fig. \ref{fig:DTikRMSE}, because it is much
larger than the shown values for LSQD. 

For the same fixed values of pixel aspect ratio and regularization strengths used for the LSQI,
Figures \ref{fig:DTikCond} and \ref{fig:DTikPix} show reconstructed image dependences on $\lambda$
and pixel aspect ratio, respectively.  As before with LSQI, the visual subjective image quality trend
in both figures aligns more with image gradient RMSE than it does with image RMSE.



\begin{figure}[t!]
	\centering
	\includegraphics{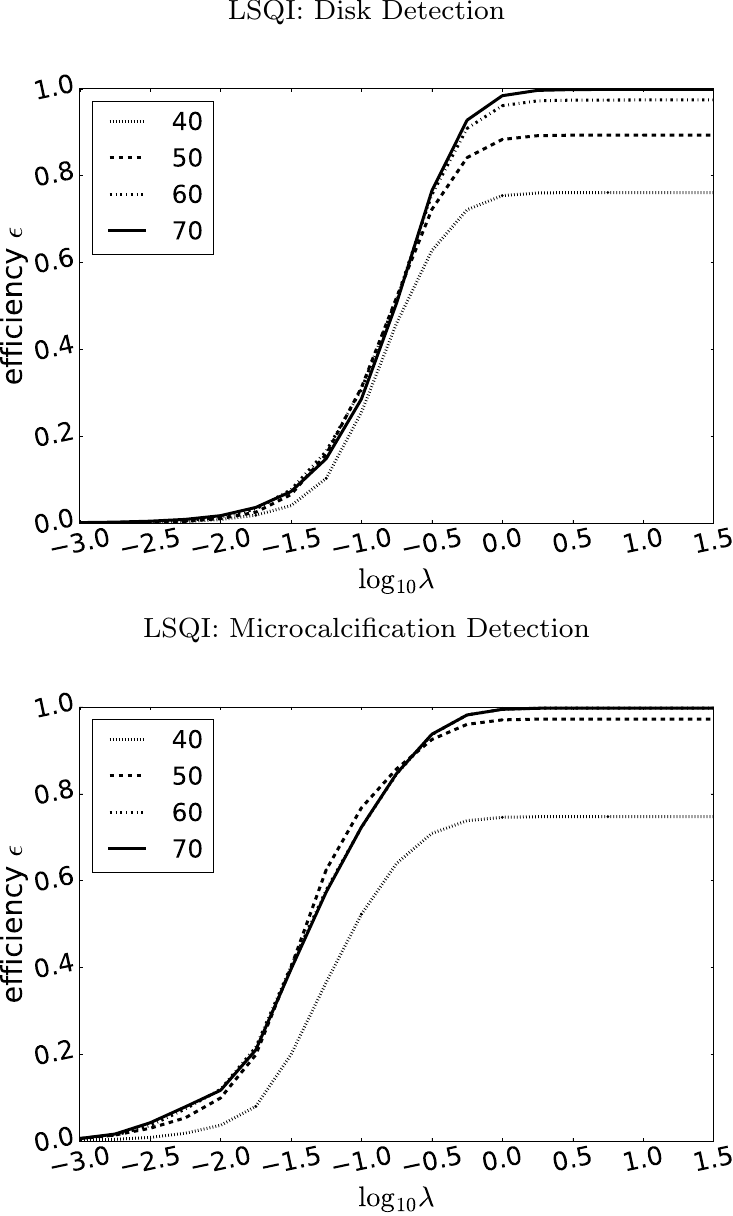}
	\caption{\label{fig:HOTik} Hotelling observer efficiency $\epsilon$ for LC disk (top) and HC calcification (bottom) 
	SKE/BKE detection tasks as a function of regularization strength
	for LSQI reconstruction. Legends indicate size of ROI (value of $l$) used for estimating the reconstruction 
	operator.}

\end{figure}


\begin{figure}[t!]
	\centering
	\includegraphics{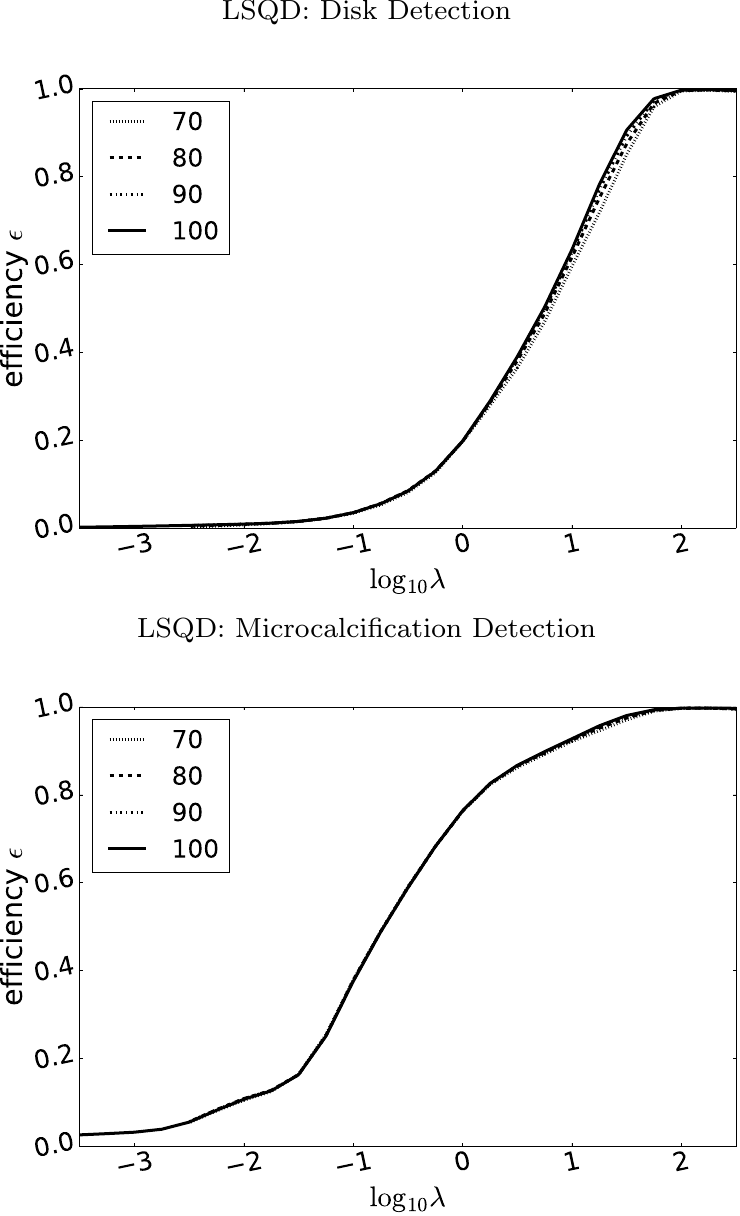}
	\caption{\label{fig:HODTik} Hotelling observer efficiency $\epsilon$ for LC disk (top) and HC calcification (bottom) 
	SKE/BKE detection tasks with variable background as a function of regularization strength
	for LSQD reconstruction.}

\end{figure}

\begin{figure}[t!]
	\centering
	\includegraphics{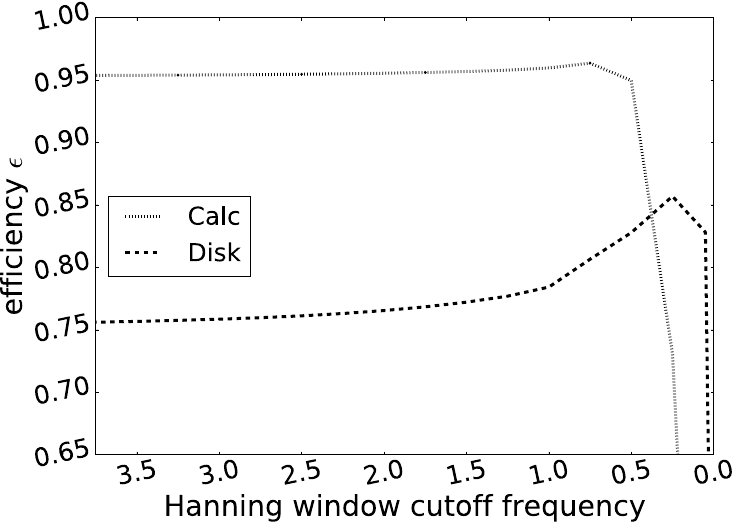}
	\caption{\label{fig:FBPEff} Hotelling observer efficiency $\epsilon$ for LC disk and HC calcification 
	SKE/BKE detection tasks as a function of spectral filter cutoff frequency
	for FBP reconstruction. Note the $x$-axis is inverted so regularization strength increases from left to right.}
\end{figure}


\subsection{Simulation study 2: ROI-HO detection efficiency as a function of regularization strength}

\subsubsection{SKE/BKE Tasks}
\textit{LSQI} - Figure \ref{fig:HOTik} shows the ROI-HO detection efficiency $\epsilon$ in LSQI reconstruction as a function of regularization strength for the two SKE/BKE detection 
tasks and 4 values of the ROI$_\text{recon}$ size parameter $l$. Recall from section \ref{sec:HOMethod} that ROI$_\text{recon}$ is
used to estimate the reconstruction operation $R$, and does not change the decimation operator
$M$, which is held constant throughout. The efficiency curves appear to approach a limiting curve as the value of $l$ is increased, 
suggesting that the largest value employed ($l=70$) provides a good approximation of the behavior of the ROI-HO. We therefore limit our discussion
to this value of $l$ for the LSQI results.

\begin{figure}[t!]
	\centering
	\includegraphics{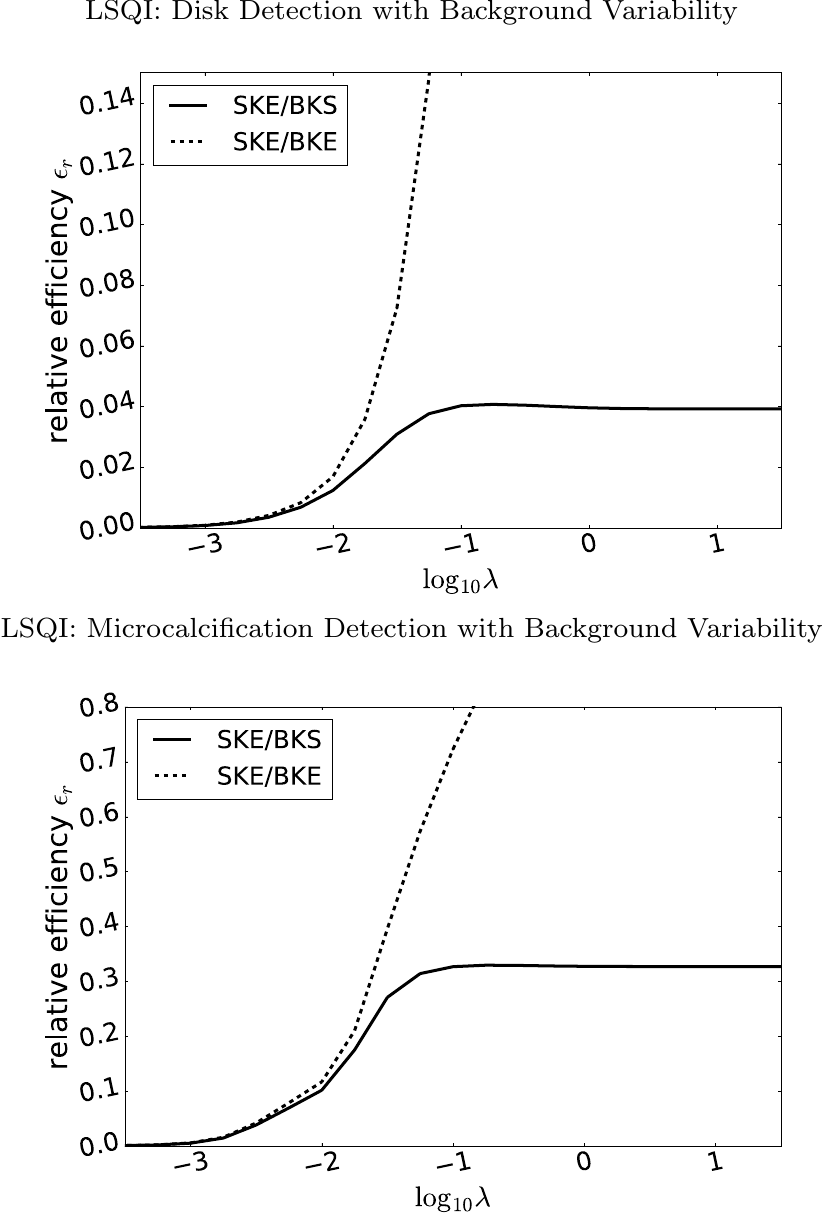}
	\caption{\label{fig:BgHOTik} Hotelling observer relative efficiency $\epsilon_r$ for LC disk (top) and HC calcification (bottom) 
	SKE/BKS detection tasks as a function of regularization strength for LSQI reconstruction. Dashed lines show HO efficiency $\epsilon$ 
	for the corresponding SKE/BKE task.}
\end{figure}

The efficiency curves are monotonically increasing with increasing regularization strength for both tasks, starting from a low-efficiency plateau, then rising and saturating at a large efficiency which is close
to the theoretical maximum of $\epsilon=1.0$. The location and width of the transition as a function
of $\lambda$ clearly depends on the characteristics of the signal.
The saturation of the efficiency curves suggests that there is a point beyond which increasing regularization
does not further improve preservation of task relevant information through the reconstruction. 
Also, there is no penalty in terms of HO SNR for detection for increasing the regularization strength
arbitrarily.
Recall from section \ref{sec:opt} that the LSQI 
reconstructed image approaches the back-projection image as $\lambda \rightarrow \infty$.

\begin{figure}[t!]
	\centering
	\includegraphics{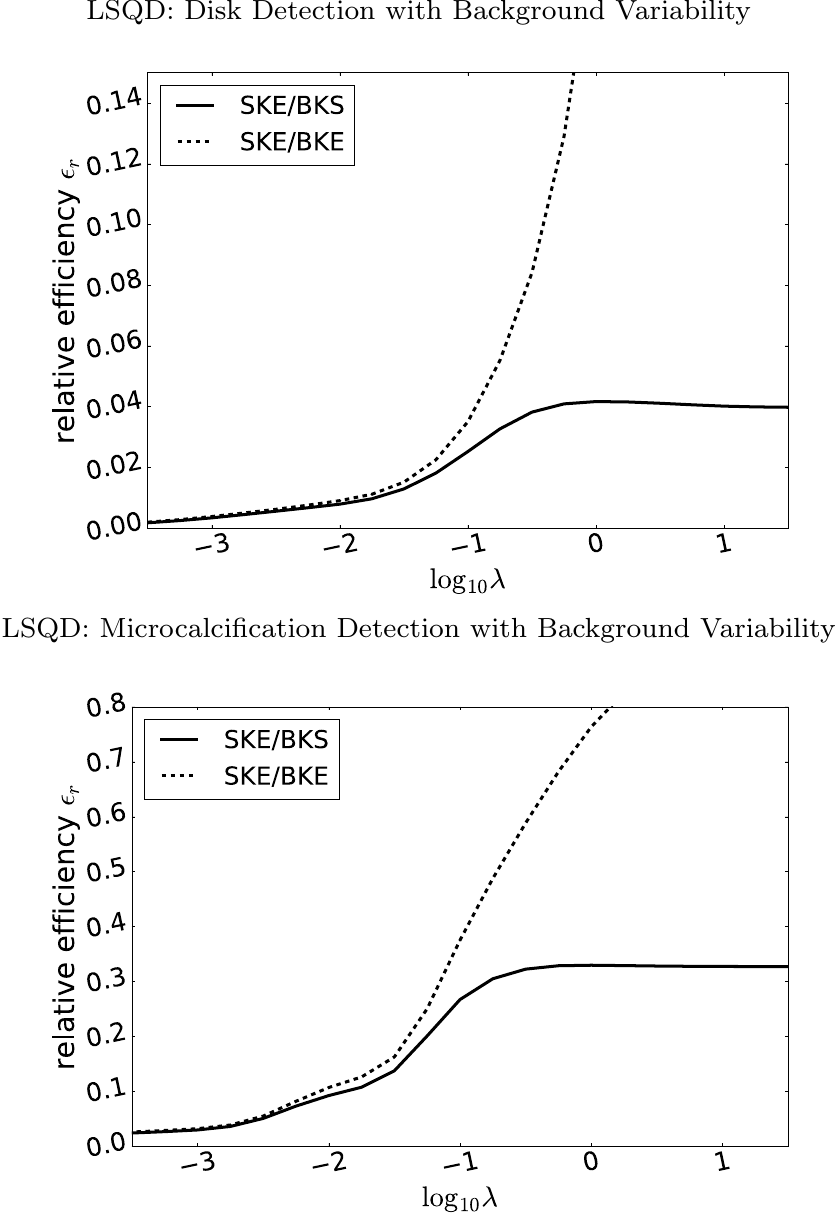}
	\caption{\label{fig:BgHODTik} Hotelling observer relative efficiency $\epsilon_r$ for LC disk (top) and HC calcification (bottom) 
	SKE/BKS detection tasks as a function of regularization strength
	for LSQD reconstruction. Dashed lines show HO efficiency $\epsilon$ 
	for the corresponding SKE/BKE task.}
\end{figure}

\textit{LSQD} - Figure \ref{fig:HODTik} shows the efficiency curves for LSQD reconstruction. The efficiency curves again approach
a limit with increasing values of $l$, in this case suggesting the value of $l = 100$ yields
a reasonable approximation of the ROI-HO. Similar to LSQI, the efficiency saturates with increasing regularization strength in a signal-dependent manner.

\textit{FBP} - Figure \ref{fig:FBPEff} shows the efficiency for FBP reconstructions of varying cutoff frequency for the Hanning apodizing window. The $x$-axis of 
the plot has been inverted so that regularization increases from left to right, consistent with the LSQI and LSQD efficiency curves.

By contrast to the LSQI and LSQD efficiency curves, the FBP efficiency curves do not display saturating behavior with increasing regularization strength, instead indicating that an SNR penalty is incurred for apodizing too heavily for a quantum noise limited task. Recall that in the limit of infinite regularization, the FBP algorithm convolves each projection with a constant function prior to back-projection.

In the limit of decreasing regularization strength, a plateau in efficiency is observed. In this limit the reconstruction algorithm approaches FBP with a pure ramp filter, so the level of the plateau is representative of the efficiency of FBP with no apodization. Moving further to the right, a monotonic increase in efficiency is observed prior to a peak. The location of the peak appears to be signal dependent, paralleling the signal-dependent location of the transition region in the efficiency curves for LSQI and LSQD. Note that in both cases, the location the of the peak/transition region for calcification detection occurs at lower regularization strength than for disk detection.


\begin{figure}[t!]
	\includegraphics{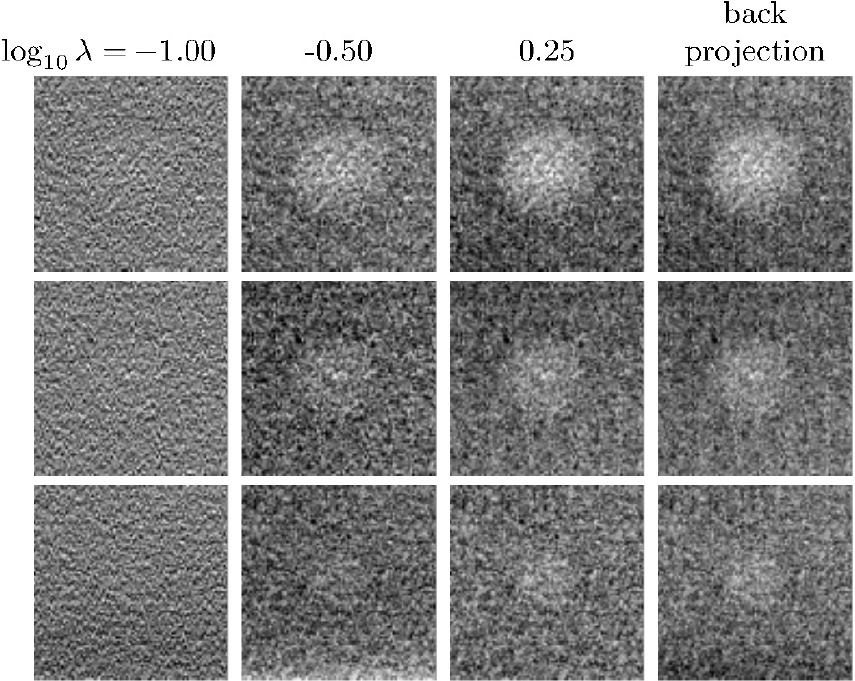}
	\caption{\label{fig:ACRTikDisk} ROIs of LSQI reconstructions containing 
	0.25, 0.5, and 0.75cm disks of ACR phantom. Regularization increases from left to right. Back projection image is shown
	for reference.}
\end{figure}
\subsubsection{SKE/BKS Tasks}
\textit{LSQI} - Figure \ref{fig:BgHOTik} shows the ROI-HO relative detection efficiency $\epsilon$ in LSQI reconstruction as a function of regularization strength for the two SKE/BKS detection 
tasks alongside the ROI-HO detection efficiencies for the corresponding SKE/BKE tasks. Based on the limiting behavior observed in the SKE/BKE tasks, only the $l=70$ ROI$_\text{recon}$ results are shown.

The relative efficiencies show a clear decrease in task performance relative to the quantum limited SKE/BKE case for both tasks, though the magnitude of the drop in SNR due to the presence of background variability is greater for the disk detection task than for the calcification detection task. Aside from this decrease in task performance, the overall trends in the efficiency curves are quite similar to their SKE/BKE counterparts. That the curves plateau at large regularization strengths indicates that the SNR is not penalized for increasing regularization strength arbitrarily, even in the presence of a variable, nonuniform background.

\begin{figure}[t!]
	\includegraphics{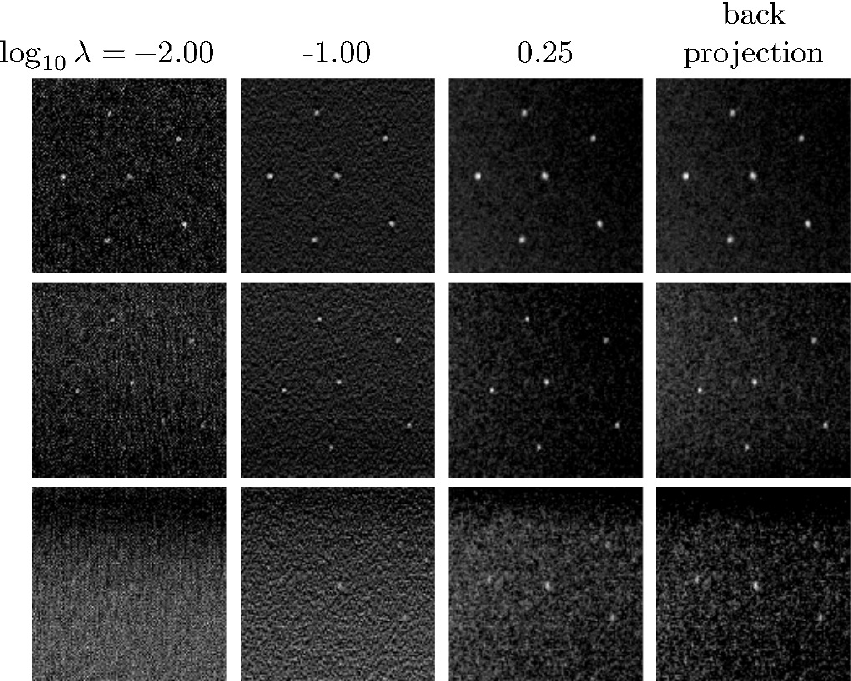}
	\caption{\label{fig:ACRTikCalc} ROIs of LSQI reconstructions containing 
	\SI{0.54}{}, \SI{0.40}, and \SI{0.32}{\milli \metre} specks of ACR phantom. Regularization increases from left to right. Back projection image is shown for reference.}
\end{figure}
At low regularization strengths, the relative efficiency curves of the SKE/BKS tasks track well with the efficiency curves of the SKE/BKE tasks. The curves separate at intermediate regularization strengths suggesting that the main contributing factor to SKE/BKS task performance at low regularization strengths is quantum noise while at large regularization strengths it is background variability. The nonuniform background variability limits the extent to which increasing regularization can improve task performance, as indicated by the lower regularization strengths at which the curves saturate for SKE/BKS tasks relative to SKE/BKE tasks.

\textit{LSQD} - Figure \ref{fig:BgHODTik} shows the corresponding results for LSQD reconstruction for the two SKE/BKS detection tasks. 
The observed behavior is largely similar to the LSQI results, with saturating behavior again occurring at large regularization strengths and a signal dependent transition region. We do note that the SNR drop for disk detection is again greater than that for calcification detection.




\subsection{Real data study: ACR phantom data}
\subsubsection{No added background}
\textit{LSQI} - ROIs of reconstructions of the ACR phantom with LSQI at three different levels of regularization, alongside a back-projection reconstruction, are
shown in Figures \ref{fig:ACRTikDisk} and \ref{fig:ACRTikCalc}. The ROIs are shown on a diagram of the ACR phantom in Figure \ref{fig:ACRDiagram} for reference.
Display windows for each reconstruction were manually chosen based on subjectively maximizing visibility of the disks or specks. We note that the
values of $\lambda$ used in the real data reconstructions may not precisely match those used in the 2D simulation studies.

As a general trend the reconstructed disks and specks appear more conspicuously as the noise-level is reduced by increasing regularization strength for the shown values of $\lambda$. Furthermore, there is good correspondence between the LSQI image
for the largest shown $\lambda$-value and the back-projection image.

\begin{figure}[t!]
	\includegraphics{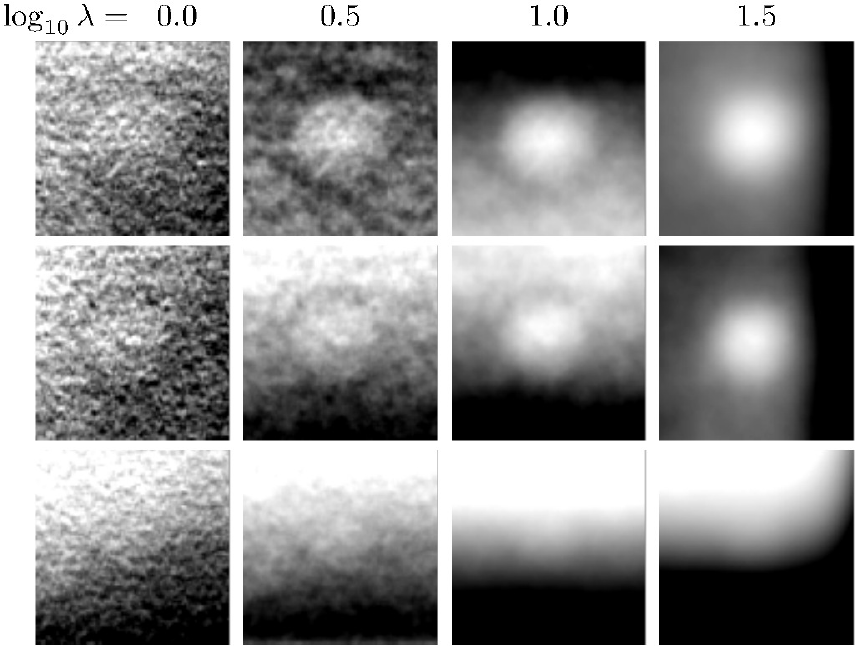}
	\caption{\label{fig:ACRDTikDisk} ROIs of LSQD reconstructions containing 
	0.25, 0.5, and 0.75cm disks of ACR phantom. Regularization increases from left to right.}
\end{figure}


The background noise texture differs significantly between the four reconstructions, with the magnitude of fluctuations
decreasing with increasing regularization strength. It is evident from the figures how the detection of large low-contrast and
small high-contrast objects could be complicated by this texture in under-regularized reconstructions.

\textit{LSQD} - Figures \ref{fig:ACRDTikDisk} and \ref{fig:ACRDTikCalc} show the analogous results
for LSQD reconstruction except that no back-projection image is shown. As with LSQI, higher levels of regularization reduce the magnitude of background noise variation;
however, there is a clear difference in noise texture. Increasing regularization of course also blurs the disk and speck signals.

\subsubsection{Added nonuniform background}
\begin{figure}[t!]
	\includegraphics{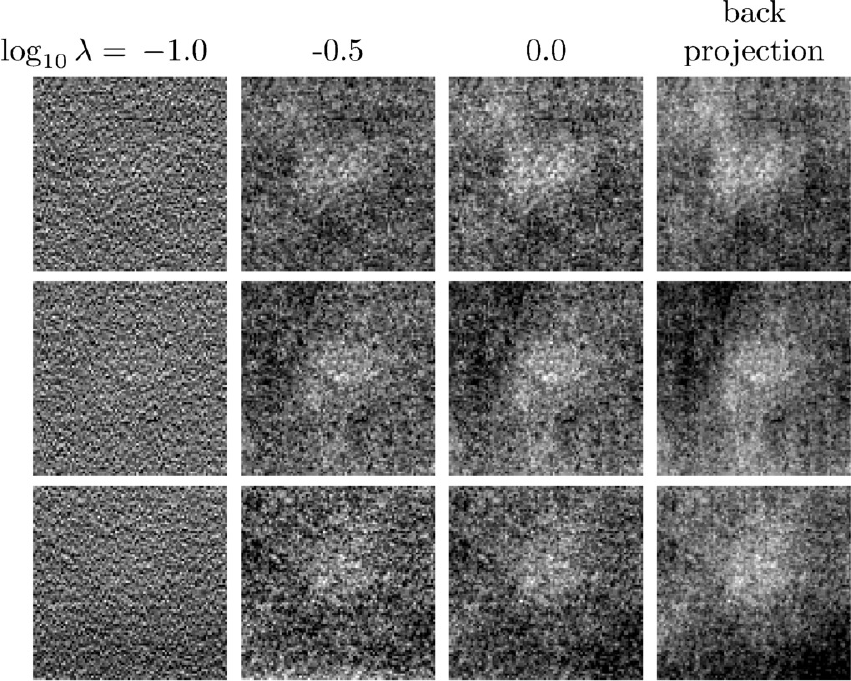}
	\caption{\label{fig:BgTikDisks} ROIs of LSQI reconstructions of altered ACR phantom data with added
	nonuniform background. ROIs contain 0.25, 0.5, and 0.75cm disks of ACR phantom. Regularization increases from left to right.
	Back projection image is shown for reference.}
\end{figure}

\begin{figure}[t!]
	\includegraphics{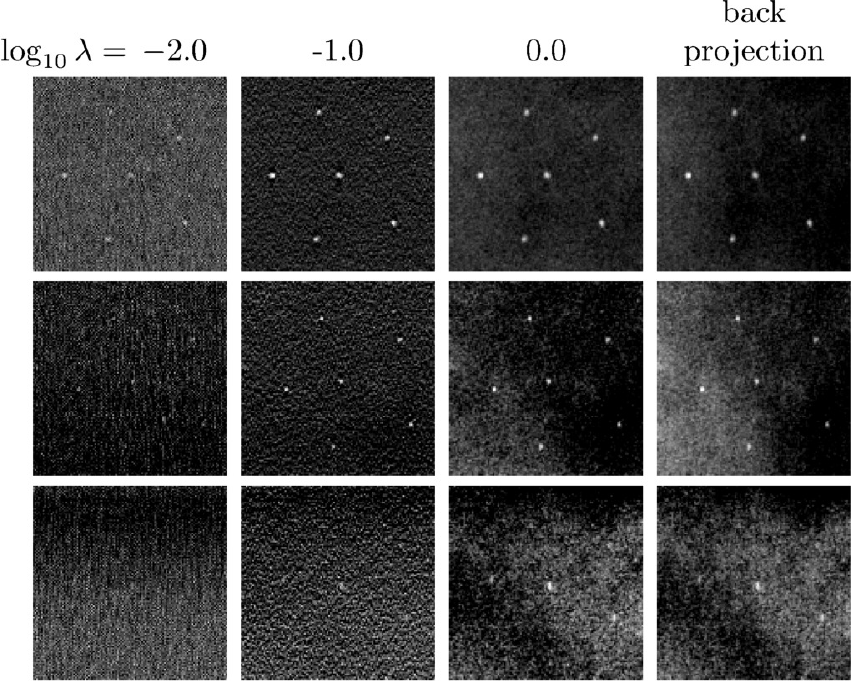}
	\caption{\label{fig:BgTikCalcs} ROIs of LSQI reconstructions of altered ACR phantom data with added
	nonuniform background. ROIs contain \SI{0.54}{}, \SI{0.40}, and \SI{0.32}{\milli \metre} specks of ACR phantom.
	Regularization increases from left to right. Back projection image is shown for reference.}
\end{figure}

\textit{LSQI} - Figures \ref{fig:BgTikDisks} and \ref{fig:BgTikCalcs} show ROIs of reconstructions of the ACR phantom in which the projection of a nonuniform background was added to the data as described in section \ref{sec:ACRmethod}. Note the the regularization parameter values for the displayed images are not identical to those shown in the corresponding results of Figures 
\ref{fig:ACRTikDisk} and \ref{fig:ACRTikCalc}.

\begin{figure}[t!]
	\includegraphics{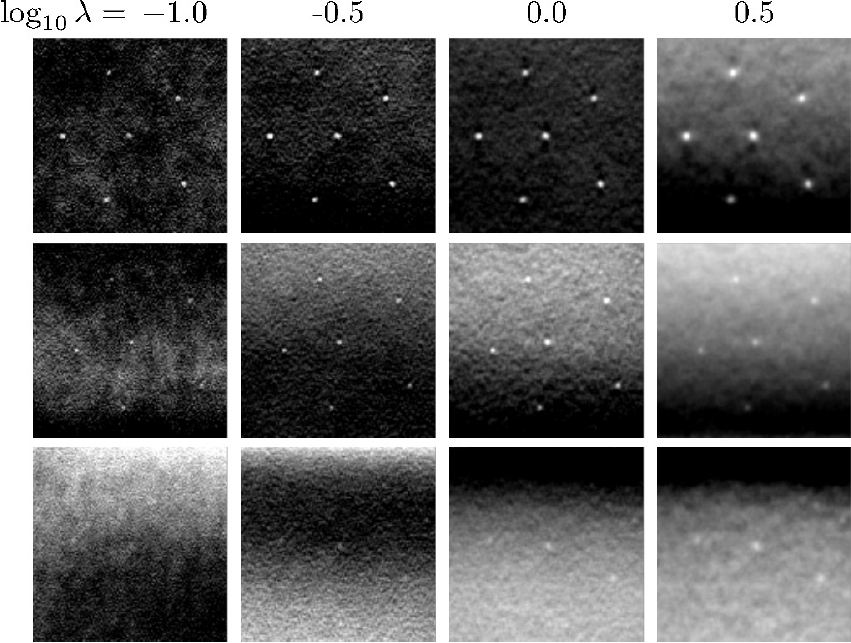}
	\caption{\label{fig:ACRDTikCalc} ROIs of LSQD reconstructions containing 
	\SI{0.54}{}, \SI{0.40}, and \SI{0.32}{\milli \metre} specks of ACR phantom. Regularization increases from left to right.}
\end{figure}

A general trend of increasing conspicuity with increasing regularization is still apparent in the presence of nonuniform background structure. 
Relative to the unaltered data reconstructions, the background texture, particularly at large regularization strengths, contains more low-frequency structure, hindering visualization of the disk and speck signals. These observations are consistent with trends observed in the ROI-HO SKE/BKS simulation studies.

\begin{figure}[t!]
	\includegraphics{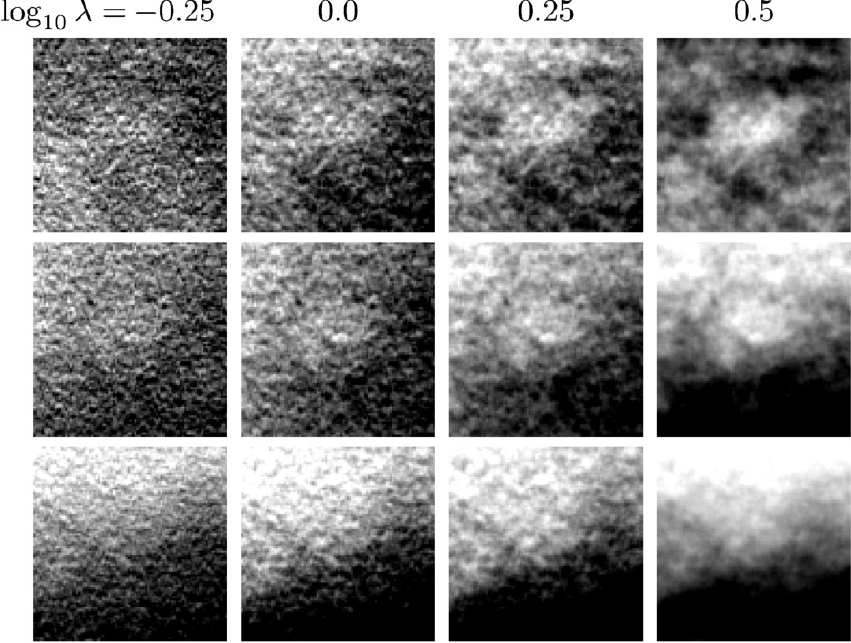}
	\caption{\label{fig:BgDTikDisks} ROIs of LSQD reconstructions of altered ACR phantom data with added
	nonuniform background. ROIs contain 0.25, 0.5, and 0.75cm disks of ACR phantom. Regularization increases from left to right.}
\end{figure}

\begin{figure}[t!]
	\includegraphics{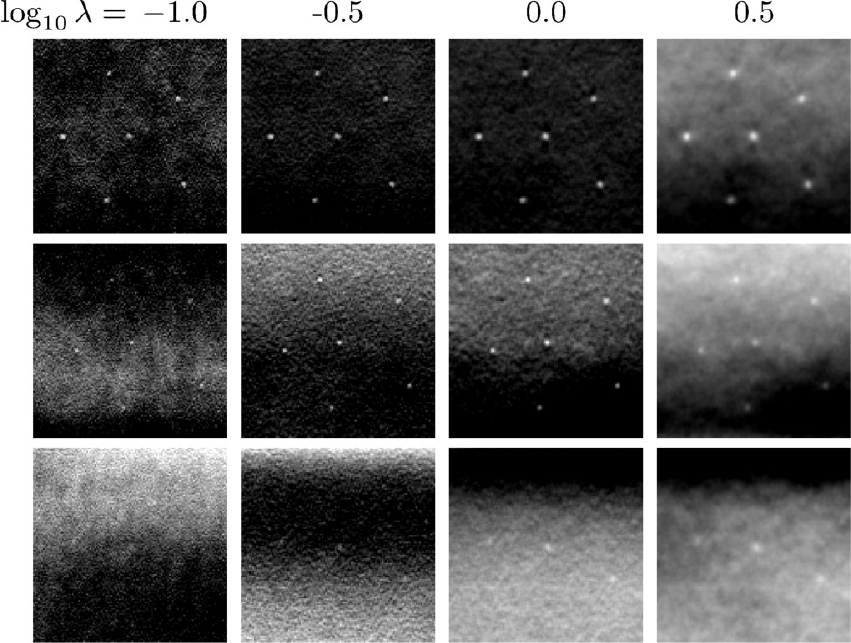}
	\caption{\label{fig:BgDTikCalcs} ROIs of LSQD reconstructions of altered ACR phantom data with added
	nonuniform background. ROIs contain \SI{0.54}{}, \SI{0.40}, and \SI{0.32}{\milli \metre} specks of ACR phantom.
	Regularization increases from left to right. }
\end{figure}

\textit{LSQD} - Figures \ref{fig:BgDTikDisks} and \ref{fig:BgDTikCalcs} show the corresponding results for LSQD reconstruction. As with LSQI, the trend of increasing conspicuity with regularization strength is maintained in the presence of a nonuniform background, though there is an apparent reduction in conspicuity of the disks relative to the unaltered data case. Conspicuity of the specks appears to be less affected by the nonuniform background (compare Figures \ref{fig:ACRDTikCalc} and \ref{fig:BgDTikCalcs}).


\section{Discussion}
\label{sec:discussion}

\textit{image RMSE} - The sensitivity of image RMSE to changes in mean pixel value
appears to make it inappropriate for meaningful characterization of image reconstruction parameters for DBT.
From the results shown the image RMSE does not follow the subjective visual quality of the reconstructed
images from the 2D DBT simulation.  Perhaps the most egregious discrepancy occurs in the image series
comparing use of various regularization strengths for LSQI, in Figure \ref{fig:TikCond}, and LSQD,
Figure \ref{fig:DTikCond}, with FBP.
The FBP result has an image RMSE which is more than twice that of any of the shown LSQI or LSDQ images,
yet it is subjectively competitive with any of these images. 

\textit{gradient image RMSE} - 
We observe that the gradient image RMSE is not as sensitive to changes in mean pixel value and appears to reflect
visual changes between different regularization strengths and aspect ratios.  In particular, an increase in gradient image RMSE
is observed at large aspect ratios for $\lambda=10^{-2.0}$, reflecting the appearance of high frequency artifacts
in both the LSQI and LSQD images shown in Figures \ref{fig:TikPix} and \ref{fig:DTikPix}, respectively. Moreover, in the image sequences varying regularization strength in Figures \ref{fig:TikCond} and \ref{fig:DTikCond}, the rank ordering suggested by the gradient image RMSE seems plausible even including the FBP images.

We caution, however, that these studies are preliminary and that many more empirical studies are needed
to support the utility of the gradient image RMSE metric for image reconstruction parameter characterization
in DBT. One peculiar aspect of this metric is that the relative magnitude of the variations of gradient image RMSE shown in Figures \ref{fig:TikRMSE} and \ref{fig:DTikRMSE} amount to 1\% to 2\% over the shown parameter ranges.
Yet, the preliminary results from subjective visualization seem to indicate that these small variations are
meaningful. That gradient image RMSE exhibits such behavior for DBT image reconstruction is perhaps not
too surprising, because the metric is still essentially comparing reconstructed images from very limited
angular-range scanning with perfect tomographic image reconstruction. By its design, the gradient image RMSE
is most sensitive to discrepancy at edge discontinuities at the borders of various tissues within the subject.
It is known that image reconstruction from a limited scanning-angular range such as in DBT can only recover
a small subset of these edges \cite{Frikel2013}.

\textit{ROI-HO efficiency for signal detection} -
The results of the second simulation study demonstrate that the proposed ROI-HO implementation can be used to accurately calculate 
ROI-HO efficiencies for the SKE/BKE and SKE/BKS detection tasks
in linear optimization-based image reconstruction for the 2D-DBT model in the scanning-arc plane. Due to non-locality of image reconstruction in general, it is not obvious that restricting the image reconstruction operator to an ROI with the image will yield accurate image domain HO SNR values.
That we observe a saturation of efficiency values (hence image domain SNR) with ROI length $\ell$ in
Figures \ref{fig:HOTik} and \ref{fig:HODTik} indicates that the ROI approximation can yield accurate
results.  We expect that ROI-HO efficiency for the 2D DBT model should reflect HO efficiency for 3D DBT,
because the 2D model captures the most important scanning geometry features of 3D DBT. This correspondence,
however, needs to be demonstrated, and we will do so in future work.

Examining the SKE/BKE task efficiency curves in more detail, both Figures \ref{fig:HOTik} and \ref{fig:HODTik}
indicate low efficiency at low regularization strength.  This efficiency loss is primarily due
to the fact that we are modeling single-slice viewing with the decimation operation that limits
the reconstruction to the line in the 2D DBT model (corresponding to a plane in 3D DBT), where
the signal for detection exists. In this line, the signal is reconstructed at an amplitude
lower than the true signal due to DBT depth blur, yet noise from the data model is amplified in the viewing
line due to low regularization.
Both LSQI and LSQD saturate at nearly a perfect efficiency of 1.0 with increasing $\lambda$, where LSQI and LSQD images limit to back-projection and the inverse Laplacian of back-projection images, respectively.

In comparing FBP and LSQI/LSQD for the SKE/BKE detection task, there are substantial differences in how the respective regularization parameters impact signal detection efficiency, highlighting the differences between these algorithms.

Turning to the SKE/BKS tasks, one might expect that with the incorporation of a nonuniform background variability model, the need for the reconstruction algorithm to resolve the signal from the background would cause a penalty to SNR at large regularization strengths for LSQI and LSQD. Contrary to this expectation, we observed monotonically increasing efficiency with increasing regularization as in the SKE/BKE tasks. The nonuniform background variability instead appeared to play the role of limiting the extent to which increasing regularization could improve task performance, while not penalizing the SNR for further increases in regularization beyond the point at which efficiency saturates. 

We also noted that the decrease in SNR relative to the SKE/BKE tasks was much greater for disk detection than calcification detection. This result suggests that the stationary process used to model nonuniform background variability interferes more with detection of large, low-contrast objects than small, high-contrast objects. This is consistent with the slowly spatially varying appearance of realizations of this background model (see, e.g., ref. \onlinecite{Burgess2007}).

In relation to other work on using the HO for image quality assessment in DBT, which
employ noise realizations \cite{VandeSompel2011, Park2010, Zeng2015, Reiser2010, Richard2010, Young2013} and/or 
assumptions of stationary noise \cite{Gang2011,Reiser2010,Richard2010} in estimating image statistics, the current approach
yields highly accurate values subject only to minor numerical error due to computer precision.
The drawback of our ROI-HO implementation is that it is based on simulation and cannot be
applied directly to reconstructed images from actual DBT systems.

\textit{3D DBT image reconstruction of the ACR phantom}-
The ACR phantom study provides a realistic visual assessment of images reconstructed in the parameter ranges investigated in the second simulation study for the ROI-HO signal detection efficiency. The viewing conditions for the unaltered data reconstructions are similar to the SKE/BKE signal detection task design. The properties
of the signals, disks and speck clusters, are known to the observer, see Figure \ref{fig:ACRDiagram};
also the background is uniform and therefore also known to the observer.
The properties of the signals of the 2D simulation study are modeled loosely on the disk and speck objects
of the ACR phantom.

The main trend seen in the unaltered ACR data reconstructions for LSQI and LSQD is that increasing regularization appears to increase the conspicuity of both disk and speck cluster signals. This trend appears to coincide
with the SKE/BKE ROI-HO efficiency trends for LSQI and LSQD. For ACR results with LSQI, in particular, the $\lambda = \infty$ images, a.k.a. back-projection, appear to show similar conspicuity for both disks and specks as the
results from the largest finite regularization strength of $\lambda= 10^{0.25}$ in Figures \ref{fig:ACRTikDisk}
and \ref{fig:ACRTikCalc}. This result is somewhat indicative of the detection efficiency plateau seen
at large $\lambda$ in Fig. \ref{fig:HOTik}. We do point out, however, that the growing magnitude of background variations relative to the blurred signal complicates
visualization at high regularization strengths for LSQD.

The apparent trend of increasing conspicuity with regularization for both disk and speck cluster signals is also observed in the ACR reconstructions in which a nonuniform background was incorporated into the data. This observation is consistent with the SKE/BKS ROI-HO efficiency trends. We note in particular that the conspicuity of the signals does not appear to be reduced at large regularization strengths, contrary to what one may expect with a nonuniform background.

We note that the correspondence between the 2D DBT ROI-HO results and 3D DBT ACR reconstructed images
represent only preliminary indications as many more empirical results are needed. Again, the connection between
the 2D DBT ROI-HO with a 3D DBT ROI-HO needs to be established. We also point out the SKE/BKS results represent a single parameter setting of a single background variability model. Other background variability models \cite{Reiser2010,Reiser2013} may yield different detection efficiency curves. Another important point is that the ROI-HO yields
image domain signal detection SNR, while for the results with the ACR phantom we only discussed signal
conspicuity. Signal conspicuity and detection, although related, are not the same thing. Closer visual correspondence with ROI-HO SNR can be investigated by designing and performing two-alternative forced choice experiments \cite{Sanchez2013}(ref. \onlinecite[p. 819]{Barrett2004}). We also note that LSQI and LSQD are not commonly used in the clinic.

\textit{Application of image quality metrics} - The metrics adapted and investigated here yield useful information
for setting parameters in linear iterative image reconstruction, but they alone do not provide a
complete picture and they are meant to compliment other image quality metrics developed for DBT systems
parameter characterization/optimization. In particular, consider the results of the ROI-HO SKE/BKE
signal detection task efficiency for LSQD and LSQI;
following the curves shown in Figures \ref{fig:HOTik} and
\ref{fig:HODTik} alone suggests use of LSQI and LSQD with extremely large regularization
parameter $\lambda$. We point out, however, that the ROI-HO signal detection efficiency exhibits
a plateau at large $\lambda$, and considering these curves together with other metrics such as the gradient image
RMSE suggests instead to employ a regularization strength at the low $\lambda$ end of the plateau.

\section{Conclusions and Future Work}
\label{sec:summary}
We have investigated three simulation-based image quality metrics for use in the selection of parameters for linear iterative image reconstruction in DBT. 
In the process, we have adapted an RMSE-based metric to DBT, and developed a task-based
metric for linear iterative image reconstruction in DBT.
The imaging properties of DBT led us to modify image RMSE to a gradient image RMSE. 
We have also extended our work on ROI-HO signal detection efficiency for filtered back-projection
\cite{Sanchez2014a} to linear iterative image reconstruction. Finally, we showed images reconstructed from
simulated 2D DBT data and ACR phantom 3D DBT data to provide a subjective visual assessment of
image reconstruction algorithm parameter trends. The results demonstrated that sensitivity to
mean pixel value is a weaknesses of the image RMSE in the context of DBT. Eliminating dependence
on mean pixel value by the proposed gradient image RMSE appears to rectify the short-comings
of image RMSE. The results also demonstrated a signal-dependent saturating behavior of the ROI-HO 
efficiency for two SKE/BKE and SKE/BKS signal detection tasks with increasing regularization strength. 

Future work will focus on extending the simulation studies from 2D to 3D in order to quantify
the correspondence of the 2D DBT model in the scanning arc plane with full 3D DBT. 
While the RMSE-based metrics can be applied to any image reconstruction algorithm including
non-linear iterative image reconstruction, the current version of the ROI-HO applies
only to linear image reconstruction. Methodology on developing noise properties for non-linear
image reconstruction, \cite{Sanchez2015,Bonetto2000} may allow extension of the ROI-HO to 
non-linear image reconstruction algorithms. 
Finally, we intend to extend the ROI-HO to more
realistic signal detection models where only statistical knowledge is available for the signal is available. Such a model is expected to be more sensitive to DBT depth resolution than
the ROI-HO for a SKE/BKE or SKE/BKS detection task. 

\section{Acknowledgements}
The authors would like to thank Hologic for providing the phantom data.
This work was supported in part by NIH R01 Grants Nos. CA158446, CA182264, EB018102, and 
NIH F31 Grant No. EB023076. The contents of this article are solely the responsibility of the authors 
and do not necessarily represent the official views of the National Institutes of Health.

\section{Disclosure of Conflicts of Interest}
The authors have no relevant conflicts of interest to disclose.

%

\end{document}